\documentclass[12pt]{article}
\usepackage{epsfig}
\usepackage{amsfonts}
\usepackage{amsmath}
\usepackage{hyperref,amssymb,amsmath,mathrsfs,bm,graphicx}
\usepackage{enumitem}
\usepackage{color}
\begin {document}

\title{Connecting the exterior gravitational field with the energy-momentum tensor of axially symmetric compact objects} 

\author{J.L. Hern\'andez-Pastora\thanks{E.T.S. Ingenier\'\i a
		Industrial de B\'ejar. Phone: +34 923 408080 Ext
		2223. Also at +34 923 294400 Ext 1574. e-mail address: jlhp@usal.es}\\
	\\
	Departamento de Matem\'atica Aplicada
	\\Instituto Universitario de F\'\i sica Fundamental y Matem\'aticas. \\ Universidad de Salamanca.  Salamanca,
	Espa\~na.  }

%\date{\today}

\maketitle

\begin{abstract}
 A method to construct interior axially symmetric metrics that appropriately match with any vacuum  solution of the  Weyl family is developed in \cite{weylsources}.  It was shown, for the case of some vacuum solutions,  that  the simplest solution for the interior metric  leads to sources with well-behaved energy conditions. Now, we integrate the field equations to obtain the  interior metric functions in terms of the anisotropies and pressures of the source.  As well, the compatible  equations of state  for these global models are calculated.  The interior metric and the suitable energy-momentum tensor describing the source are constructed in terms of the exterior metric functions. At the boundary of the compact object,  the behaviour of a pressure $T_m$, defined from the energy-momentum tensor, is shown to be related with the exterior gravitational field. This fact allows us to explore the differences arising at the matter distribution when the spherical symmetry of the global metric is dropped. Finally, an equation derived from the matching conditions is obtained which allows us to calculate the Weyl coefficients of the exterior metric as source integrals. Hence the Relativistic Multipole Moments of the global model can be expresed in terms of the matter distribution of the source.
\end{abstract}

Pacs numbers:04.20.Cv, 04.20.Dw, 97.60.Lf, 04.80.Cc,,  04.20.Cv, 04.20.-q, 4.20.Ha, 95.30.Sf.

\maketitle

\section{Introduction}

The Einstein's field equations on the inside of any compact object establish a relation between the matter content of the source and the interior metric, which is referred to as the interior solution; whereas the vacuum field equations provide us with the suitable exterior metrics, compatible with the symmetry of the problem and  matching appropriately with the interior metric, which  allows us to describe the gravitational field of the source.
The aim of this work consists on  establishing, beyond the requirements of the matching conditions,  a precise relation between  the exterior metric and the  interior solution, or in other words the gravitational field and the energy-momentun tensor of the source. I want to focus on the exterior solution to construct from it the global metric as well as the suitable matter contents of the source.
 
 Starting from some particular interior solution for a relativistic source describing a compact stellar object, its exterior gravitational field is unique, and it is given by a solution of the  Einstein's vacuum field equations. If we restrict ourselves to the case of axially symmetric metrics then that solution belongs to the Weyl family \cite{Weyl}. The detailed analysis of this exterior metric provides a large amount of information about the source; this is the aim of many research papers devoted to describe the physics of point mass particles in the presence of the gravitational field. For example the study of the gyroscopic precession \cite{gyros} allows to distinguish  gravitational fields, as well as the study of circular orbits \cite{circular}, gravitational radiation \cite{RG} or collapse processes \cite{collapse}. Another fruitfull technique is based on the use of the Relativistic Multipole Moments (RMM) \cite{paperrmm}, \cite{thorne} 
 to describe the gravitational field within their gravitational effects on the test particles. 
 
On the contrary, and unlike the vacuum case,  the Einstein's fied equations on the inside of the source show a much richer scenario, in the following sense: if one starts with a known exterior metric and try to obtain a source, then a lot of possible solutions, not only for the interior metric but for different matter distributions, can be managed. The most common procedure to face this fact leading to obtain global models of self graviting compact objects, defines a working  direction from the inside to the outside. It is to say, the process of solving starts in the  interior field equations compatibles with different matter distributions and/or any symmetry considerations and culminates with the suitable exterior Weyl metric matching appropriately at the boundary of the object.

 This dynamic of work was recently reversed in  \cite{weylsources}, \cite{kerrsources} where authors calculate interior metrics successfully matching with either any of the Weyl family vacuum solutions or stationary axially solutions (in particular the Kerr solution). The relevance of this result becomes highlighted  in view especially of the fact that the interior line element  is constructed in terms of the exterior metric functions evaluated at the boundary, in such a way that both the metric and the energy-momentum tensor are related with the gravitational field. In addition, the RMM arise as physical parameters related with the source. This is the aim of the work in \cite{RMMsource} where the RMM are defined as quantities calculated from volume integrals trying to generalize the Newtonian Multipole moments defined in Classical Gravity. Another interesting result is \cite{gurle} where volume integral espressions are obtained to construct the RMM.

 In \cite{weylsources} it was shown that, for some Weyl exterior solutions (specifically the Zipoy-Vorhees \cite{zipoy} and the Monopole-Quadrupole solutions \cite{MQ}, \cite{MQ1}) the interior metric obtained with the minimum requeriments of the matching restrictions derived from the procedure (i.e., by taking the functions $\mathbb{F}=\mathbb{G}=0$, see \cite{weylsources} for details) lead to suitable configurations of the energy-momentum tensor $T_{\mu \nu}$ with a good physical behaviour. 
 
 In the work that I am introducing now we are able to integrate the Einstein's  axially symmetric field equations providing the interior metric functions in terms of the matter content. The matching conditions along with the field equation themselves impose  a set of restrictions on the energy-momentum tensor $T_{\mu \nu}$. Therefore, each one of the exterior gravitational field  delimits/ reduces/ constrains /restricts the suitable behaviours of the matter distribution (pressures, anisotropies and equation of state EoS). Global solutions to the gravitation of a compact object connect both the interior and the vacuum metric with the source, and the more relevant feature of this fact is, in my opinion, that it allows us to relate the exterior gravitational field with the energy-momentun tensor $T_{\mu \nu}$ of the compact object.

 It is not true  that any $T_{\mu \nu}$ is able to integrate the field equations providing a succesfull metric inside the source, but indeeed  a suitable matching with some exterior Weyl solution is even more  difficult to obtain. The result that we are introducing now  is the possibility of  knowledge from the gravitational field about the physics of the source by means of the behaviour that the physical parameters of that object must  satisfy to be considered the source of that field.
 This result provides more information about the source beyond the symmetries of the problem and it constrains the physics of the object that is generating such gravitational field. In particular the RMM structure of the source is connected with those quantities defined within the exterior metric \cite{thorne}. A collateral conclusion of this procedure leads to get the set of the  Weyl coefficients by means of volume integrals over the source, which allows us to construct the RMM in terms of the matter distribution.

To achieve the main goal of this work, we shall extensively use the general method to construct global static axially symmetric solutions to Einstein equations deployed in \cite{weylsources}. A very brief rewiew of this method is presented in the next section, all the details may be found in  that reference\footnote{As we shall explain in the next section some misprints are found in that paper}. The aim of this paper is to integrate the field equations in order to obtain the interior metric functions in terms of the energy-momentum tensor of the source. Solutions for  spherical as well as non-spherical sources are found, and they are related with the anisotropies of the sources. Compatible equations of state are determined.

%%%%%%%%%%%%%%%%%%%%%%%%%%%
%%%%%%%%%%%%%%%%%%%%%%%%%%%%

\section{The global static and axisymmetric metric in the Erez-Rosen coordinates}

We shall write the  global static and axisymmetric line element in the 
Erez-Rosen coordinates:
\begin{equation}
ds^2=-e^{2\sigma} dt^2+e^{2\nu} dr^2+e^{2 \eta} r^2 d\theta^2+e^{2 \mu} r^2\sin^2\theta d\varphi^2,
\label{ERglobal}
\end{equation}
where the metric functions, which  depend on $r$ and $\theta$, are defined as follows
\begin{equation}
e^{2\sigma}=\left\lbrace
\begin{matrix}
& e^{2\hat a} Z^2 &  \ , \ r\leq r_{\Sigma}, & \frac{e^{2\hat g-2\hat a}}{A}  \nonumber \\
\nonumber \\
& e^{2 \psi} & \ , \ r\geq r_{\Sigma}, & \quad e^{-2\psi+2 \hat{\gamma}}\nonumber
\end{matrix}
\right\rbrace=e^{2\nu}
\end{equation}
\begin{equation}
e^{2\eta}=\left\lbrace
\begin{matrix}
& \frac{e^{2\hat g-2\hat a}}{A}  &  \ , \ r\leq r_{\Sigma},& \qquad  e^{-2\hat a} \nonumber \\
\nonumber \\
& e^{-2\hat{\psi}+2 \hat{\gamma}}& \ , \ r\geq r_{\Sigma}, & \qquad e^{-2\hat{\psi}}
\end{matrix}
\right\rbrace=e^{2\mu}
\end{equation}

$r=r_{\Sigma}$ being the boundary surface of the source for a constant value $r_{\Sigma}$, $M$ denotes the mass in the Schwarzchild space-time, $\displaystyle{A\equiv 1-\frac{2Mr^2}{r_{\Sigma}^3}}$,  $Z=\frac 32 \sqrt{A( r_{\Sigma})}-\frac 12 \sqrt{A}$ and $\psi$, $\gamma$ are any metric functions of the Weyl family of vacuum solutions, $\hat{\gamma}\equiv \gamma-\gamma_s$, $\hat{\psi}\equiv \psi-\psi_s$, $\gamma_s$, $\psi_s$ being the metric functions corresponding to the Schwarzschild solution, and $\hat a$, $\hat g$  are suitable functions constructed in \cite{weylsources} in order to guarantee a good physical behaviour of the energy-momentum tensor and the matching (Darmois) conditions \cite{darmois}.

The interior metric function $\hat a=a-a_s$ is constructed in such a way that $a(r_{\Sigma})=\psi(r_{\Sigma})$, $a_s(r_{\Sigma})=\psi_s(r_{\Sigma})$, also  $\hat g=g-g_s$ is such that $g(r_{\Sigma})=\gamma(r_{\Sigma})$, $g_s(r_{\Sigma})=\gamma_s(r_{\Sigma})$, together with the condition that the corresponding derivatives with respect to the radial variable of both $\hat a$ and $\hat g$ equal the derivatives of the exterior metric functions $\psi$, $\gamma$ at the boundary.

The general solution for the exterior metric function is given by the Weyl family in the Erez-Rosen representation \cite{erezrosen}
\begin{equation}
\psi=\sum_{n=0}^{\infty}(-1)^{n+1} q_n Q_n\left(\frac{r}{M}-1\right) P_n(\cos\theta),
\label{erezrosenfamily}
\end{equation}
where $P_n(\cos \theta)$ are Legendre Polynomials, $Q_n(y)$ are Legendre functions of second kind and $q_n$ a set of
arbitrary constants.  The relationship between the canonical Weyl coordinates $\lbrace R, \omega\rbrace$, and the Erez-Rosen $\lbrace r, y=\cos\theta \rbrace$ system is as follows
\begin{eqnarray}
R&=&\sqrt{(r-M)^2-M^2(1-y^2)},\nonumber \\
\omega&=&y\frac{( r -M)}{R}.
\label{weylcoord}
\end{eqnarray}

In addition, to guarantee a good behaviour of the physical variables at the center of the matter  distribution we shall demand:
\begin{equation}
\hat a_0=\hat a^{\prime}_0=\hat g_0=\hat g^{\prime}_0=\hat g^{\prime \prime}_0=0,
\label{condcero}
\end{equation}
where prime denotes derivative with respect to  $r$, and  the subscript $0$ indicates that the quantity is evaluated at the origin of coordinates.
Then all the above conditions lead to the following expressions for the interior metric functions (the subscript $\Sigma$ denotes that the function is evaluated at $r=r_{\Sigma}$)
\begin{eqnarray}
\hat a(r,\theta)&=&\hat \psi_{\Sigma} s^2(3-2s)   +r_{\Sigma}\hat \psi^{\prime}_{\Sigma}s^2(s-1)+\mathbb{F}(r,\theta),\nonumber \\
\hat g(r,\theta)&=&\hat \gamma_{\Sigma} s^3(4-3s)   +r_{\Sigma}\hat \gamma^{\prime}_{\Sigma}s^3(s-1)+\mathbb{G}(r,\theta).
\label{aygsimple}
\end{eqnarray}
with $s\equiv r/r_{\Sigma} \in \left[0,1\right]$ and  $\mathbb{F},\mathbb{G}$ are arbitrary functions with the following behaviour at both extremes of the interval, i.e., at the center and the boundary:
\begin{eqnarray}  &\mathbb{F}(s=0,1,\theta)=\mathbb{F}^{\prime}(s=0,1,\theta)=0,\nonumber\\ &\mathbb{G}(s=0,1,\theta)=\mathbb{G}^{\prime}(s=0,1,\theta)=0, \quad  \mathbb{G}^{\prime\prime}(0,\theta)=0.
\label{boundarycond}
\end{eqnarray}

These metric functions, satisfy the junction conditions   and generate physical variables which are regular within the fluid distribution.  Furthermore  the vanishing of  $\hat g$ on the axis of symmetry, as required by the regularity conditions,  necessary to ensure elementary flatness in the vicinity of  the axis of symmetry, and in particular at the center,
is assured by the fact that $\hat \gamma_{\Sigma}$ and $\hat \gamma^{\prime}_{\Sigma}$ vanish on the axis of symmetry.

Even more, at this level of generality we can assure that the junction conditions  imply the vanishing of the radial pressure $(P_{rr}\equiv g_{rr}T^1_1)_{\Sigma}=0$ at the boundary, and it can be shown that  $T_1^2$ vanishes on the boundary surface as well \cite{weylsources}.

\vskip 5mm

For the special case $\hat a=\hat g=0$ we recover the spherical perfect fluid with isotropic pressures scenario:

\begin{eqnarray}
ds^2_I&=& -Z^2 dt^2+ \frac{1}{A( r)} d{ r}^2+{ r}^2 (d \theta^2+\sin^2\theta d\varphi^2),\nonumber \\
ds^2_E&=& -\left(1-\frac{2M}{r}\right) dt^2+ \frac{1}{1-\frac{2M}{r}} d{ r}^2+{r}^2( d \theta^2+\sin^2\theta d\varphi^2).
\label{sphericalER}
\end{eqnarray}

Thus, the global line element (\ref{ERglobal}) describes in the vacuum any solution of the Weyl family ($\psi, \gamma$) and a good behaved  interior solution with an isotropic perfect fluid limit when the matching is performed  with the Schwarzschild space--time.
In \cite{weylsources} the case $\mathbb{F}=\mathbb{G}=0$ was studied for some examples, in particular the resulting sources for the exterior field of the MQ$^{1}$  \cite{MQ1} and Zipoy-Vorhees \cite{zipoy}--\cite{9} solutions.

 Now, the point is that, for any exterior gravitational field an infinite number of sources exist. Accordingly the obvious questions arises: how can we restrict the possible sources of a given exterior solution belonging to the Weyl family?, Can we use  our knowledge of the  exterior gravitational field to constrain the energy-momentum tensor $T_{\mu \nu}$?

In what follows we shall see how to answer to the above questions by integrating the field equations in order to obtain the interior metric functions in terms of the tensor $T_{\mu \nu}$. It may be used to limit the possible sources for any Weyl solution. The resulting constraints may be interpreted in terms of specific restrictions on the equation of state of the source and its anisotropies. At the same time, once we set a physical configuration of the source compatible with the field equations, the geometry of the source is described in terms of that physics of the source and the exterior gravitational field.

%%%%%%%%%%%%%%%%%%%%%%%%%%%%%%%%%%
%%%%%%%%%%%%%%%%%%%%%%%%%%%%%%%%%%

\section{The field equations}

From the line element (\ref{ERglobal}), the non-vanishing components of the  energy momentum tensor are related with the interior metric functions as follows (see \cite{weylsources} for details\footnote{Please take into account  a missprint in that paper for the expression of  $E$ and $\hat p_{zz}$, as well as in the formula (24) in that paper derived from the previous mistaken formulae: the second derivative of the function $\hat g$ with respect to the variable $s$ must contain a forgotten  factor $A$. Same missprints are reproduced in \cite{kerrsources}. The calculations and conclusions derived in both papers are still appropriated and right, since it is a matter of a missprint in the edition of the  latex version.} ):
\begin{eqnarray}
-T^0_0&=&\kappa \left(8 \pi \mu+\hat p_{zz}-E\right),\nonumber\\
T^1_1&=& \kappa \left(8 \pi P-\hat p_{xx}\right),\nonumber \\
T^2_2&=& \kappa \left(8 \pi P+\hat p_{xx}\right),\nonumber \\
T^3_3&=&\kappa \left(8 \pi P-\hat p_{zz}\right), \nonumber \\
T_1^2&=&g^{\theta\theta}T_{12}=- \frac{\kappa }{r^2} \ \hat p_{xy} ,
\label{eegeneral}
\end{eqnarray}
where the following notation has been used
$\displaystyle{\kappa\equiv \frac{e^{2\hat a-2\hat g}}{8 \pi}}$, superscripts $(\dot{\ })$ and $(\ddot{\ })$  denote first and second derivatives respectively with respect to  angular variable $\theta$, and $\Delta$ denotes the laplacian operator with axial symmetry:  

$$\displaystyle{\Delta \hat a= \hat a^{\prime \prime}+2\frac{\hat a^{\prime}}{r}+\frac{\ddot{\hat a}}{r^2}+\frac{\dot{\hat a}}{r^2}\frac{\cos \theta}{\sin \theta}}$$

\begin{eqnarray}
E&=&-2 \Delta \hat a+(1-A)\left[2 \frac{\hat a^{\prime}}{r}\frac{9 \sqrt{A_{\Sigma}}-4 \sqrt{A}}{3 \sqrt{A_{\Sigma}}- \sqrt{A}}+2 \hat a^{\prime \prime}\right],\nonumber\\
\hat p_{xx}&=&-\frac{\dot{\hat a}^2}{r^2}-\frac{\hat g^{\prime}}{r}+\hat a^{\prime 2}+\frac{\dot{\hat g}}{r^2}\frac{\cos \theta}{\sin \theta}+\nonumber\\
&+&(1-A)\left[2 \frac{\hat a^{\prime}}{r}\frac{\sqrt{A}}{3 \sqrt{A_{\Sigma}}- \sqrt{A}}- \hat a^{\prime 2} +\frac{\hat g^{\prime}}{r}\frac{3 \sqrt{A_{\Sigma}}-2 \sqrt{A}}{3 \sqrt{A_{\Sigma}}-\sqrt A}\right], \nonumber \\
\hat p_{zz}&=&-\frac{\dot{\hat a}^2}{r^2}-\frac{\hat g^{\prime}}{r}-\hat a^{\prime 2}-\frac{\ddot{\hat g}}{r^2}-\hat g^{\prime \prime}+\nonumber\\
&+&(1-A)\left[-2 \frac{\hat a^{\prime}}{r}\frac{\sqrt{A}}{3 \sqrt{A_{\Sigma}}- \sqrt{A}}+ \hat a^{\prime 2} +2\frac{\hat g^{\prime}}{r}+\hat g^{\prime \prime}\right], \nonumber\\
\hat p_{xy}&=&2 \dot{\hat a}  \hat a^{\prime}-\hat g^{\prime}\frac{\cos\theta}{\sin\theta}-\frac{\dot{\hat g}}{r}+\frac{(1-A)}{r \sqrt A (3 \sqrt{A_{\Sigma}}- \sqrt{A})}(2 \dot{\hat a} -\dot{\hat g} )
\label{eegeneraldet}
\end{eqnarray}

In the general axially simmetric case there are three independent pressures, as can be deduced from the eigenvalues  $\lambda$ of the tensor $T_{\mu \nu}$:
\begin{eqnarray}
\lambda_0&=&T^0_0=-\kappa \left(8 \pi \mu+\hat p_{zz}-E\right),\nonumber\\
\lambda_3&=&T^3_3=\kappa \left(8 \pi P-\hat p_{zz}\right), \nonumber \\
\lambda_{\pm}&=&\frac 12 \left[T^1_1+T^2_2 \pm\sqrt{(T_1^1-T_2^2)^2+4T_1^2 T_2^1} \right]= \nonumber\\
&=&\kappa \left(8 \pi P\pm \sqrt{\hat p_{xx}^2+\hat p_{xy}^2\frac{A}{r^2}}\right),
\label{autovalores}
\end{eqnarray}
where $\lambda_0$ is related to the density and $\lambda_3$, $\lambda_{\pm}$ represent the three pressures. From above (\ref{autovalores}) and by taking into account that  $\lambda_{\pm}=\lambda_3+k\Pi_{\pm}$, with $\Pi_{\pm}\equiv\hat p_{zz}\pm \sqrt{\hat p_{xx}^2+\hat p_{xy}^2\frac{A}{r^2}}$ then we can introduce the three independent anisotropies: $\left(\kappa \Pi_{\pm},2 \sqrt{\hat p_{xx}^2+\hat p_{xy}^2\frac{A}{r^2}}\right)$. Nevertheless, it is preferable to redefine the anisotropies  we will work with onwards by rewritting the field equations in terms of them as follows:
\begin{eqnarray}
\Pi_{31} 8\pi\frac{P}{T_m}&=&\hat p_{xx}-\hat p_{zz} \nonumber\\
\Pi_{23} 8\pi\frac{P}{T_m}&=&\hat p_{xx}+\hat p_{zz} \nonumber\\
\Pi_{xy} 8\pi\frac{P}{T_m}&=&-\frac{1}{r^2}\hat p_{xy}\nonumber\\
-T_0^0 8\pi\frac{P}{T_m}&=&8\pi \mu+\hat p_{zz}-E
\label{anisotroEC}
\end{eqnarray}
where the notation $T_m\equiv\frac{T_1^1+T_2^2}{2}$ is used and  $\Pi_{ij}\equiv T_i^i-T_j^j$, $\Pi_{xy}\equiv T_1^2$. The above four equations  (\ref{anisotroEC}) together with the following relation obtained from the sum  $T_1^1+T_2^2$:
\begin{equation}
\frac{T_1^1+T_2^2}{2}\equiv T_m=8\pi \kappa P
\label{11mas22}
\end{equation}
become into the new five field equations, equivalent to the set (\ref{eegeneral}), which will be used from now onwards. The last one of the equations  	(\ref{anisotroEC}) provides the equation of state (EoS) and the other ones the three independent anisotropies $\Pi_{31}, \Pi_{23},\Pi_{xy}$. Let us note that the anisotropy $\Pi_{21}=T_2^2-T_1^1$ is not independent one since it is a linear combination of the two first equations  (\ref{anisotroEC}). With respect to the sum of the other components of the tensor $T_{\mu}^{\nu}$ we have $T_1^1+T_3^3=2 T_m-\Pi_{23}$ and $T_2^2+T_3^3=2 T_m+\Pi_{31}$.
	
	When the particular spherical case is regarded then $\hat a=\hat a(r), \hat g=\hat g(r)$, and the following equation is obtained from (\ref{eegeneral}) 
	\begin{equation}
	8 \pi r^2 T_1^2e^{2\hat g-2\hat a}=\hat g^{\prime} \frac{\cos \theta}{\sin\theta} \rightarrow T_1^2=t(r)\frac{\cos \theta}{\sin\theta}
	\label{t12esfe}
	\end{equation}
	where ${\displaystyle t(r)\equiv\frac{\hat g^{\prime}}{8\pi r^2}e^{2\hat g-2\hat a}}$. Therefore, we must take  $t(r)=0$ if we want to avoid infinite pressures at the symmetry axis ($\theta=0$), or in other words we must consider a vanishing component $T_1^2$ of the energy-momentum tensor. Hence, the spherical case implies $\hat g^{\prime}=0$, which together with the matching conditions (\ref{boundarycond}) is equivalent to  $\hat g=0$. In addition, for this case we have that  $\hat p_{xx}=-\hat p_{zz}$ and then  $T_2^2=T_3^3$, which means that only two independent pressures exist in this case namely  $T_1^1$ and $T_2^2$. Those pressures are used to be denoted in literature  as $p_r$ (radial pressure) and  $p_{\bot}$ ( tangential pressure) whenever the spherical gauge is used for the coordinates (those coordinates preserve the angular part of the metric in the form $ r^2 d\theta^2+r^2\sin^2\theta d\phi^2$). In such system of coordinates we have that  $T_1^2=0$, $T_1^1=8\pi p_r$ y $T_2^2=T_3^3=8\pi p_{\bot}$. 
	
	In conclusion, no matter the system of coordinates we were using, the spherical case leads to a unique independent anisotropy $p_r-p_{\bot}$, which corresponds to $\Pi_{31}$ in our notation (since $\Pi_{31}=\Pi_{21}$ and $\Pi_{23}=0$) with the following field equation:
	\begin{equation}
	\Pi_{31}8\pi \frac{P}{T_m}=2 \hat p_{xx}.
	\label{31esferico}
	\end{equation}

\section{The solutions for the metric functions}

	We proceed now to integrate the equations (\ref{anisotroEC}, \ref{11mas22}). Let us note that the equation (\ref{11mas22}) allows us to establish a relation between both interior metric functions as follows:
	\begin{equation}
	\hat g=\hat a-\frac 12 \ln\left(\frac{T_m}{P}\right)
	\label{gvsa}
	\end{equation}
	
	\subsection{Isotropic and anisotropic spherical sources}
	
	All the spherical sources only match with the Schwarzschild exterior metric. Since $\hat a=\hat a(r)$, and we have constructed the metric functions as follows (\ref{aygsimple})
	\begin{equation}
	\hat a=a_{E}\left(\hat \psi_{\Sigma}(\theta),\hat \psi_{\Sigma}^{\prime}(\theta);r\right)+\mathbb{F}, \ \hat g=g_{E}\left(\hat \gamma_{\Sigma}(\theta),\hat \gamma_{\Sigma}^{\prime}(\theta);r\right)+\mathbb{G}
	\end{equation}
then, $a_E=g_E=0$ and $\hat a=\mathbb{F}(r)$. Since both $T_1^2$ and $\hat g$ vanish as we have explained before, the remaining field equations (\ref{anisotroEC}, \ref{11mas22}) are  obtained from (\ref{gvsa}) and (\ref{31esferico}) as follows:	
\begin{equation}
\mathbb{F}=\hat a=\frac 12 \ln\left(\frac{T_m}{P}\right)\label{efesfe}
\end{equation}
\begin{equation}
\Pi_{31}8\pi \frac{P}{T_m}=2 \hat p_{xx}=2\hat a^{\prime}\left(A \hat a^{\prime}+\frac{2(1-A)}{r} \frac{\sqrt{A}}{3\sqrt{A_{\Sigma}}-\sqrt{A}}\right).
\label{31pxx}	
\end{equation}
and along with  the last equation of (\ref{anisotroEC}) (EoS) that we shall analyze in the next section.

	On the one hand, the isotropic perfect fluid case requires $\hat p_{xx}=-\hat p_{zz}=0$ leading to $T_1^1=T_2^2=T_3^3$. The only possible solution satisfying at the same time the junction conditions, is $\hat a=\mathbb{F}=0$, and therefore  we get from (\ref{31pxx})  the vanishing of the anisotropy $\Pi_{31}=0$. Hence, the isotropic perfect fluid limit stands for $T_m=P$.
	
	On the other hand, the anisotropic case establishes the following relation between pressures $P$, $T_m$ and the anisotropy $\Pi_{31}$ (\ref{31pxx}):
\begin{equation}
\Pi_{31}=A\frac{T_m}{P}\left( \frac{T_m^{\prime}}{T_m}- \frac{P^{\prime}}{P}\right)\left[\frac 14 \left( \frac{T_m^{\prime}}{T_m}- \frac{P^{\prime}}{P}\right)-\frac{P^{\prime}}{\mu+P} \right]
\end{equation}

\subsection{Non-spherical sources}
	
From the previous   first three field equations  (\ref{anisotroEC})	 we obtain, by taking into account the equation (\ref{gvsa}), the following set of non-linear partial differential equations involving the metric function  $\hat a$:

	\begin{eqnarray}
&\hat a^{\prime 2}+\hat a^{\prime} \left(\frac{1-A}{2Ar}\frac{4\sqrt{A}-3\sqrt{A_{\Sigma}}}{2Z}\right)+\frac 12 \hat a^{\prime\prime}+\frac{1}{2Ar^2}\left(\frac{\cos\theta}{\sin\theta}\dot{\hat a}+\ddot{\hat a}\right)+d(r,\theta)=0 \nonumber\\
& \dot{\hat a}^2+\hat a^{\prime} \left(rA+\frac{r(1-A)}{2}\frac{2\sqrt{A}-3\sqrt{A_{\Sigma}}}{2Z}\right)
+\frac {Ar^2}{2} \hat a^{\prime \prime}+\frac{1}{2}\left(-\frac{\cos\theta}{\sin\theta}\dot{\hat a}+\ddot{\hat a}\right)+\hat d(r,\theta)=0   \nonumber\\
& \hat a^{\prime}\dot{\hat a}+\dot{\hat a}\left( \frac{1-3\sqrt{A A_{\Sigma}}}{4rZ\sqrt{A}}\right)-\frac{\cos\theta}{2\sin\theta}\hat  a^{\prime}=\alpha(r,\theta)
\label{123}
\end{eqnarray}
where the following notation has been used
	\begin{eqnarray}
	d(r,\theta) &=&-\frac{4\pi}{A} \Pi_{31} \frac{P}{T_m}+\frac{3\sqrt{A_{\Sigma}}(1-A)}{8ArZ}\left( \frac{T_m^{\prime}}{T_m}- \frac{P^{\prime}}{P}\right)+\frac{\xi}{2Ar^2} \frac{\partial_{\theta}T_m}{T_m}+ \nonumber\\
	&-&\frac{1}{4} \left[ \frac{T_m^{\prime\prime}}{T_m}-\frac{P^{\prime\prime}}{P}-\frac{T_m^{\prime2}}{T_m^2}+\frac{P^{\prime2}}{P^2}\right]-\frac{1}{4 Ar^2}  \left[ \frac{\ddot{T_m}}{T_m}-\frac{\dot{T_m}^2}{T_m^2}\right] \nonumber \\
	\hat {d}(r,\theta) &=& 4\pi r^2 \Pi_{23} \frac{P}{T_m}+r\frac{-4AZ+(1-A)(3\sqrt{A_{\Sigma}}-2\sqrt{A})}{4Z}\left( \frac{T_m^{\prime}}{T_m}- \frac{P^{\prime}}{P}\right)+ \nonumber\\
	&-&\frac{\xi}{2} \frac{\partial_{\theta}T_m}{T_m}-\frac{Ar^2}{4} \left[ \frac{T_m^{\prime\prime}}{T_m}-\frac{P^{\prime\prime}}{P}-\frac{T_m^{\prime2}}{T_m^2}+\frac{P^{\prime2}}{P^2}\right]-\frac 14  \left[ \frac{\ddot{T_m}}{T_m}-\frac{\dot{T_m}^2}{T_m^2}\right]\nonumber\\
	\alpha(r,\theta)&=&-8\pi r^2\Pi_{xy} \frac{P}{T_m}+\frac 12 \xi\left( \frac{T_m^{\prime}}{T_m}- \frac{P^{\prime}}{P}\right)-\frac{1-2A-3\sqrt{A_{\Sigma}A}}{8r\sqrt{A}Z}\frac{\partial_{\theta}T_m}{T_m}
	\label{alfayotras}
	\end{eqnarray}

Now we take the first  of the equations (\ref{123}) multiplied by a factor $Ar^2$ and substract the second of those equations, and  then an equivalent system of equations   can be obtained as follows:

	\begin{equation}
	 {h^{\prime}}^2+h^{\prime} \left( \frac{(1-A)\sqrt{A}-2Z(1-3A)}{4ZrA}\right)+\frac 12 h^{\prime\prime} +\frac{1}{2Ar^2}\left(\frac{\cos\theta}{\sin\theta}\dot{h}+\ddot{h}\right)+\beta=-d(r,\theta)
	\label{latercera}
	\end{equation}
		\begin{eqnarray}
	& A r^2 {h^{\prime}}^2-\dot{h}^2=\Lambda  \nonumber\\
	&\dot{h} h^{\prime}=\Omega
	\label{123eny}
	\end{eqnarray}
	with the notation
	\begin{eqnarray}
	\Lambda&=& \delta-\xi^2+Ar^2x^2 \nonumber\\
	\Omega&=& \alpha+x \xi \nonumber\\
	\delta&=& 4\pi r^2(\Pi_{31}+\Pi_{23}) \frac{P}{T_m}-\frac{r \sqrt{A}(1-2A+3\sqrt{A_{\Sigma}A})}{4Z}\left( \frac{T_m^{\prime}}{T_m}- \frac{P^{\prime}}{P}\right)-\xi \frac{\partial_{\theta}T_m}{T_m} \nonumber\\
	\beta&=&-\frac{(1-A)^2+(2-A)(2Z)^2}{4r^2A(2Z)^2}
	\label{lambdaomegadeltaybeta}
	\end{eqnarray}
	
	and the partial derivatives of the metric function  $\hat a$ have been redefined by means of introducing a new function $h$ as follows:
	\begin{eqnarray}
	h^{\prime}&=&\hat a^{\prime}+x(r), \qquad x(r)\equiv \frac{1-3\sqrt{A A_{\Sigma}}}{4rZ\sqrt{A}} \nonumber\\
	\dot{h}&=&\dot{\hat a}+\xi(\theta) , \qquad \xi(\theta)\equiv -\frac 12 \frac{\cos\theta}{\sin\theta}
	\label{redefi}
	\end{eqnarray}
	This fact is possible because $x=x(r)$ and $\xi=\xi(\theta)$, and hence the integrability condition for both  $\hat a$ and $h$ remains unchanged.

{\bf A)} One way to obtain a solution of the system of equations   (\ref{123eny}) consists on extracting  the partial derivative $\dot h$ from the second equation and putting it into the first one, leading to the following  bicuadratic equation for the partial derivative $h^{\prime}$ as follows:
\begin{equation}
Ar^2{h^{\prime}}^4-\Lambda {h^{\prime}}^2-\Omega^2=0
\label{bicuadratica}
\end{equation}
whose four admissible solutions are
	\begin{equation}
	h^{\prime}= \frac{\pm 1}{r \sqrt{2A}}\sqrt{\Lambda \pm \sqrt{\Lambda^2+4Ar^2\Omega^2}}
	\label{bicuadraticasol4}
	\end{equation}
	From all this set of solutions only one of them is suitable for our purposes since the sign inside the root in (\ref{bicuadraticasol4}) must be taken positive to preserve a real solution because  $\Lambda^2+4Ar^2\Omega^2>\Lambda^2$ ($A>0$). And with respect to the global sign in (\ref{bicuadraticasol4}) it must be taken negative to satisfy the boundary conditions  as we shall see in the Appendix. Also the integrability condition for the function $h$ shall be discussed therein, once  the system of equations (\ref{123eny}) becomes the following system:
	\begin{eqnarray}
	h^{\prime}&=&H \nonumber\\
	\dot{h}&=&\frac{\Omega}{H}
	\label{cuadratura1}
	\end{eqnarray}
where $H$ is the real and compatible solution of the bicuadratic equation (\ref{bicuadraticasol4}) considered  as follows
	\begin{equation}
	H\equiv\frac{-1}{r \sqrt{2A}}\sqrt{\Lambda + \sqrt{\Lambda^2+4Ar^2\Omega^2}}
	\label{H}
	\end{equation}

{\bf B)} Another way of writting the solution of the system of equations  (\ref{123eny}) comes from the complex equation obtained by adding the first equation of that system to the second one multiplied by the complex factor  $i\ 2r\sqrt{A}$:
	\begin{equation}
r\sqrt{A} h^{\prime}+i\dot{h} =\sqrt{\Lambda+2 i r\sqrt{A} \Omega},
\label{complejo}
\end{equation}	
in such a way that the real and imaginary parts of the two complex roots of the right-hand side of equation (\ref{complejo}) can be identified  with the partial derivatives $r\sqrt{A} h^{\prime}$ and $\dot{h}$ respectively, i.e.,
\begin{eqnarray}
	h^{\prime}&=& \pm \frac{S^{1/4}}{r\sqrt{A}}\cos(\phi) \nonumber\\
	\dot{h}&=&\pm S^{1/4}\sin(\phi)
	\label{cuadratura2}
\end{eqnarray}
where ${\displaystyle S\equiv \Lambda^2+4Ar^2\Omega^2}$, and ${\displaystyle \phi\equiv \frac 12 \arctan(u) }$ with ${\displaystyle u\equiv \frac{2r\sqrt{A}\Omega}{\Lambda}}$. These expressions in  (\ref{cuadratura2})  exactly recover the equation  (\ref{cuadratura1}) if the following trigonometric relations are considered 
\begin{equation}
\cos(u/2)=\frac{1}{\sqrt{2}}\sqrt{1+\frac{1}{\sqrt{1+\tan^2(u)}}}, 
\end{equation}
\begin{equation}
\sin(u/2)=\frac{1}{\sqrt{2}}\sqrt{1-\frac{1}{\sqrt{1+\tan^2(u)}}}
\label{cosenotg}
\end{equation}
	
	\vskip 4mm

With respect to the equation  ({\ref{latercera}) it becomes a partial differential equation for the functions $\Lambda$ and $\Omega$ once we put into it the solution (\ref{H}). Nevertheless, the resolution of that equation together with the integrability condition of (\ref{cuadratura1}) is an extremely difficult task to do. For convenience, we can consider both the equation (\ref{latercera}) and the integrability condition equation (\ref{ic3}) as partial differential equations for two independent functions, namely  $\Sigma_{\pm} \equiv \sqrt{\pm\Lambda+\sqrt{S}}$ instead of $\Lambda$ and $\Omega$, as follows:
	
	\begin{equation}
\dot{\Sigma}_{+}=r\sqrt{A}\Sigma_{-}^{\prime}
\label{icSs}
\end{equation}
	\begin{equation}
-2\sqrt{2}A r^2 \hat{\beta}-\left[\sqrt{2}\Sigma_{+}^2+\Sigma_{+}\left(\frac{-1+2A-3\sqrt{A A_{\Sigma}}}{2Z} \right)-r\sqrt{A}\Sigma_{+}^{\prime}\right]=2 \xi \  \Sigma_{-}-\dot{\Sigma}_{-}
\label{22Ss}
\end{equation}

where $\hat{\beta}\equiv\beta+d(r,\theta)$.
Once a solution for $\Sigma_{\pm}$ has been obtained, then $\Lambda$ and $\Omega$ are calculated  in terms of it. Let us  use $\sqrt{v_{\pm}}$ to denote particular solutions for $\Sigma_{\pm}$ respectively, and therefore $v_{+}+v_{-}=2\sqrt{S}$ and $v_{+}-v_{-}=2\Lambda$. Consequently,
\begin{eqnarray}
\Lambda&=& \frac 12 (v_{+}-v_{-}) \nonumber \\
\Omega&=& \frac{\sqrt{v_{+}v_{-}}}{2r\sqrt{A}} 
\label{uves}
\end{eqnarray}
We proceed to solve the pair of above equations (\ref{icSs}) and (\ref{22Ss}) by means of the achivement of $\Sigma_{-}$ in terms of the other function $\Sigma_{+}$. The consistency of that system of equations (\ref{icSs})-(\ref{22Ss})  holds iff the following condition is satisfied by $\Sigma_{+}$: 
\begin{equation}
-\ddot{\Sigma}_{+}+2 \xi \dot{\Sigma}_{+}=r\sqrt{A} \partial_r \Gamma
\label{consistency}
\end{equation}
	with the notation 
	
			$${\displaystyle \varGamma\equiv-2\sqrt{2}A r^2 \hat{\beta}-\left[\sqrt{2}\Sigma_{+}^2+\Sigma_{+}\left(\frac{-1+2A-3\sqrt{A A_{\Sigma}}}{2Z} \right)-r\sqrt{A}\Sigma_{+}^{\prime}\right]}$$
		
Therefore we need to solve the equation (\ref{consistency}) for $\Sigma_{+}$   with the appropriated boundary conditions (\ref{Bcero}), (\ref{uncuartoequiv}), and afterwards the function $\Sigma_{-}$
is obtained as the  general solution of the system of equations (\ref{icSs})-(\ref{22Ss}) as follows 
\begin{equation}
\Sigma_{-}=\int\frac{\dot{\Sigma}_{+}}{r \sqrt{A}} \ dr
\label{sigmaprima}
\end{equation}

In the section VI we shall show a suitable solution for the function $\Sigma_{+}$ and we obtain from the above  general equations  a sample model for the interior metric functions, anisotropies and pressure compatible with the exterior gravitational field.

\vskip 2mm

Finally, once  both the integrability condition  ${\displaystyle \partial_{\theta}(h^{\prime})=\partial_r(\dot{h})}$ and the equation (\ref{latercera}) have been impossed on the function $H$, then  the metric function $\mathbb{F}$ is integrated as follows
\begin{eqnarray}
\mathbb{F}&=&\int \mathbb{F}^{\prime} \ dr =-a_E+\int (-x+h^{\prime}) \ dr=\nonumber \\
&=&-a_E+\frac 12\ln\left( \frac{r}{2Z}\right)+\int H \ dr
\label{efeintegrada}
\end{eqnarray}
since  $\mathbb{F}^{\prime}= -a_{E}^{\prime}-x+h^{\prime}$.	The explicite equation (\ref{ic3}) derived from the integrability condition  is obtained in the Appendix.

\section{Constraints and equations of state}	

\subsection{The spherical case}

In this particular case, we  are only constrained by the boundary conditions regarding the metric function  $\mathbb{F}$ (\ref{boundarycond}), which implies that (see equation  (\ref{efesfe})) the pressure $T_m$ behaves just like  $P$ at the extremes of the interval $[0,r_{\Sigma}]$, i.e.,
	\begin{eqnarray}
T_m(s=0)=P(s=0)&=&\mu\left( \frac{1-\sqrt{A_{\Sigma}}}{3\sqrt{A_{\Sigma}}-1}\right) \nonumber \\ T_m(s=1)=P(s=1)&=&0
\label{efesphericextremos}
\end{eqnarray}
The same conclusion is derived for the respective derivatives because the derivative of equation  (\ref{efesfe}) and the previous conclusion  (\ref{efesphericextremos}) lead to
	\begin{eqnarray}
T_m^{\prime}(s=0)&=&P^{\prime}(s=0)=0 \nonumber \\ T_m^{\prime}(s=1)&=&P^{\prime}(s=1)=-\frac{M \mu}{r_{\Sigma}(r_{\Sigma}-2 M)}
\label{deriefesphericextremos}
\end{eqnarray}

The equation of state (EoS)  for this case  can be obtained by putting the solution (\ref{efesfe}) and $\hat g=0$ into the fourth equation (\ref{anisotroEC}), in such a way that 
\begin{equation}
-T_0^0 8\pi=\frac{T_m}{P}\left[8\pi \mu -\hat a^{\prime}\left( \frac 2r(1-3A)+A\hat a^{\prime}\right)+2A\hat a^{\prime \prime}\right]
\end{equation}
and hence the EoS proves to be

\begin{eqnarray}
\rho\equiv-T_0^0&=&\frac{T_m}{P}\left[ \mu-\frac{1-3A}{8\pi r}\left(\frac{T_m^{\prime}}{T_m}-\frac{P^{\prime}}{P} \right)+\frac{A}{8\pi}\left(\frac{T_m^{\prime\prime}}
{T_m}-\frac{P^{\prime\prime}}{P} \right)+\right.\nonumber\\
&+&\frac{A}{16\pi}\frac{T_m^{\prime}}
{T_m}\frac{P^{\prime}}{P}+\left. \frac{A}{32\pi} \left( 3\left(\frac{P^{\prime}}{P}\right)^2-5\left(\frac{T_m^{\prime}}{T_m}\right)^2\right)\right]
\label{eosesf}
\end{eqnarray}

Let us note that the EoS of the  isotropic case, $\rho=\mu$, is recovered if we take  $T_m=P$ not only at the extremes but everywhere in $[0,r_{\Sigma}]$.
	
\subsection{The non-spherical case}	
	
In the general axially symmetric case we have to take into account the boundary conditions for  both $\mathbb{F}$ and $\mathbb{G}$, and the  requirements of the integrability condition in (\ref{cuadratura1}) or (\ref{cuadratura2}). Let us address the details of the analysis to the Appendix,  since we obtain a great amount of information about that,  and we summarize here the results obtained  therein:

The pressure $T_m$ compatible with the solution obtained for the metric functions has to verify the following behaviour at the boundary $r_{\Sigma}$ (see eqs.(\ref{Fprima})-(\ref{TmprimasenE}) in Appendix): 
\begin{equation}
 T_m(s=1)=P(s=1)=0
\label{TmE}
\end{equation}
\begin{equation}
 T_m^{\prime}(s=1)=-\frac{M\mu}{r_{\Sigma}(r_{\Sigma}-2M)} e^{2\hat \Psi_{\Sigma}-2\hat\gamma_{\Sigma}}
\label{TmprimaE}
\end{equation}
\begin{equation}
 T_m^{\prime\prime}(s=1)=-\frac{M\mu e^{2\hat \Psi_{\Sigma}-2\hat\gamma_{\Sigma}}}{r_{\Sigma}(r_{\Sigma}-2M)} \left[\frac{1}{r_{\Sigma}}+4\left(\hat \Psi^{\prime}_{\Sigma}-\hat \gamma^{\prime}_{\Sigma} \right) \right].
\label{Tmprima2E}
\end{equation}
\begin{equation}
\dot T_m(s=1)=0 ,\qquad  \frac {\dot T_m}{T_m}(s=1)=2\dot{\hat \psi}_{\Sigma}-2\dot {\hat \gamma}_{\Sigma}
 \label{TmdotE}
 \end{equation}
 \begin{equation}
 \dot T_m^{\prime}(s=1)=-(2\dot{\hat \psi}_{\Sigma}-2\dot {\hat \gamma}_{\Sigma})\frac{M\mu}{r_{\Sigma}(r_{\Sigma}-2M)} e^{2\hat \Psi_{\Sigma}-2\hat\gamma_{\Sigma}}
 \label{TmdotprimaE}
 \end{equation}
As can be seen these relevant expressions allow us to relate the  derivatives of the pressure $T_m$  (at the surface of the compact object) with the exterior gravitational field. Besides, those expressions  show the different behavior arising for  the energy-momentum tensor when non-spherical symmetry is considered from outside, i.e., only if we consider Schwarzschild metric as the exterior gravitational field then we can see from (\ref{TmprimaE}) and (\ref{Tmprima2E}) that pressure $T_m$ behaves in the same way as $P$ in the boundary and we recover the expressions of the spherical case (\ref{deriefesphericextremos}).

With respect to the behaviour of $\mathbb{F}$ and $\mathbb{G}$ at the origin, it was required in \cite{weylsources} to obtain  components of the energy-momentum tensor without divergencies at the center of the compact object. But in this work we are calculating the metric functions in terms of those components of the tensor and hence the conditions on the origin can be disregarded. Nevertheless,  at the Appendix we have applied these conditions to the  functions $\mathbb{F}$ and $\mathbb{G}$ leading to the following results (from eq. (\ref{Fprima})):
\begin{equation}
 T_m(s=0)=P(s=0)=\mu \frac{1-\sqrt{A_{\Sigma}}}{3\sqrt{A_{\Sigma}}-1} ,
 \label{cero1}
 \end{equation}
 \begin{equation}
 T^{\prime}_m(s=0)=P^{\prime}(s=0)=0
 \label{cero2}
 \end{equation}
 \begin{equation}
 \partial_{\theta}T_m(s=0)=O(r^2)=0,
\label{Tmdotcero}
\end{equation}
\begin{equation}
\left[\partial_r^2\left(\frac{\partial_{\theta}T_m}{T_m} \right)\right]_0= 
 \frac{4\pi}{\xi}(\Pi_{31}+\Pi_{23})|_0  \  \frac{3\sqrt{A_{\Sigma}}-1}{3\sqrt{A_{\Sigma}}}
\label{cero3}
\end{equation}

Some comments must be outlined about these results. Firstly, by constrast with the behaviour at the boundary, the pressure $T_m$ at the origin does not depend on the exterior gravitational field. Whatever the exterior metric has been matched with the source, $T_m$ behaves appropriately at the origin in either non-spherical case or in the spherical scenario, and the unique difference between both scenarios arises at the derivative $\partial_{\theta}T_m$ (\ref{cero3}). Equations (\ref{Tmdotcero}) and (\ref{cero3}) hold that the angular derivative of the pressure $T_m$ evaluated at the origin is an infinitesimal of second order and the coefficient of that order is related with  the value at the origin of the anistropies $\Pi_{31}+\Pi_{23}$, i.e.,
\begin{equation}
{\dot T_m}^{\prime\prime}(s=0)=\frac{4\pi  \mu}{ \xi} \  \frac{1-\sqrt{A_{\Sigma}}}{3\sqrt{A_{\Sigma}}} (\Pi_{31}+\Pi_{23})|_0
\label{dthetadosprima}
\end{equation}

Secondly, let us note  that the above equations (\ref{cero1}), (\ref{cero2}) and  (\ref{Tmdotcero}) do not imply any restriction on the anisotropies since the functions $\Lambda$ and $\Omega$ evaluated on both  extremes of the interval $[0,r_{\Sigma}]$ do not involve them as can be seen from the equations (\ref{lambdaomegadeltaybeta}), (\ref{alfayotras}).  
 Hence, on despite of the above conclusion (\ref{dthetadosprima}), the anisotropies are free to verify the equation derived from the integrability condition of (\ref{cuadratura1}) which is evaluated at the Appendix leading to the equivalent equations (\ref{ic1}), (\ref{ic2}), (\ref{ic3}). These equations  can not be solved in the general case, but a solution of them is always possible since the functions  $\Lambda$ and $\Omega$ depend arbitrarily on three independent functions $\Pi_{31}$, $\Pi_{23}$, $\Pi_{xy}$. Hence, the integrability condition and the equation (\ref{latercera}) establish a relation between the anisotropies and the pressure $T_m$. In the following section we shall  ilustrate these results with a simple model by means of a particular solution of the whole system of equations (\ref{latercera}), (\ref{123eny}).

\vskip 2mm

In addition, we obtain an equation that allows us to calculate the Weyl coefficients (in the Erez-Rosen representation) $q_n$ of the exterior gravitational field (\ref{erezrosenfamily}) as an integral over the source. Since the RMM (Relativistic Multipole Moments) are constracted as certain combinations of those Weyl coefficients, this result seems to be suitable in order to establish a relation between the RMM and the source. Another authors have been working in this issue (see \cite{gurle} and references therein, \cite{RMMsource}).
We can integrate the equation (\ref{efeceroy1}) in the angular variable as we show now  and use the known expression (\ref{erezrosenfamily}) for the metric function $\psi$ in Erez-Rosen coordinates (assuming equatorial symmetry) to obtain

\begin{equation}
\int_{-1}^{1} \hat \psi_{\Sigma}P_{2k}(y) dy=\int_0^{r_{\Sigma}}dr\int_{-1}^{1}dy (-x+h^{\prime})P_{2k}(y) =-\sum_{n=1}^{\infty} q_{2n} Q_{2n}|_{\Sigma}
\frac{2 \delta_{2k,2n}}{4n+1}
\label{mmr1}
\end{equation}

where the orthogonality condition of the Legendre polynomials $P_{2n}(y)$ has been used and $Q_{2n}|_{\Sigma}\equiv Q_{2n}(r_{\Sigma}/M-1)$ denote the Legendre functions of second kind evaluated at the boundary. Taking into account that $x$ does not depend on the angular variable and hence its angular integration vanishes we have for $k>0$  that 
\begin{equation}
-q_{2k} Q_{2k}\left(\frac{r_{\Sigma}}{M}-1\right)\frac{2}{4k+1}=\int_0^{r_{\Sigma}}dr\int_{-1}^{1}dy \  P_{2k}(y) H
\label{mmr2}
\end{equation}
Henceforth we can evaluate the Weyl coefficients as integrals over the source. Since we know the explicit expressions of the RRM in terms of those coefficients \cite{tesis} we are able to obtain integral expressions  extended over the source leading to connect those quantities with the interior extructure of the source. For example the quadrupole moment $Q$ of any axially simmetric solution with equatorial symmetry is known \cite{tesis}, \cite{MQ} to be ${\displaystyle Q=\frac 13a_0^3-a_2}$, $a_i$ being the Weyl coefficients in Weyl coordinates or equivalently ${\displaystyle Q=\frac{2}{15} q_2}$, $q_i$ being the coefficient of the Erez-Rosen family \cite{tesis}, \cite{erezrosen} of axially symmetric solutions, since the relation between both sets of parameters is known \cite{tesis}: ${\displaystyle a_0=-M, a_2= -M^3\left(\frac 13 q_0+\frac{2}{15} q_2\right)}$ (where $q_0$ is taken to be equal to one for solutions of the Erez-Rosen family with spherical limit of Schwarzschild). Therefore, from (\ref{mmr2}) the relativistic quadrupole moment can be obtained by the folowing integral equation
\begin{equation}
Q=-\frac{1}{3 Q_2 |_{\Sigma}} \int_0^{r_{\Sigma}}dr\int_{-1}^{1}dy \  P_{2}(y) H
\label{Q}
\end{equation}
with ${\displaystyle Q_2 |_{\Sigma}=\frac{3(r_{\Sigma}-M)^2-M^2}{4M^2} \ln(A_{\Sigma})-\frac 32\frac{r_{\Sigma}-M}{M} }$.

\vskip 2mm

Let us note the relevance of this integral equation since it shows the  contribution of the internal sctructure of the source, in particular the anisotropies and the pressure $T_m$ (involved in $H$) to conform the quadrupole moment of the solution.

\vskip 2mm

Finally, we address the issue of the EoS for these global models. 
From the last equation of (\ref{anisotroEC}) and by using the relation  between the metric functions (\ref{gvsa}) we conclude that

\begin{equation}
\rho\equiv -T_0^0 =\frac{T_m}{ 8\pi P}\left[8 \pi \mu+ -\frac{\dot {\hat a}^2}{r^2}+\frac{2\dot {\hat a}}{r^2}\frac{\cos\theta}{\sin\theta}+\frac{\ddot {\hat a}}{r^2}+\frac{{\hat a}^{\prime}}{r}(-1+4A)+A\hat a^{\prime \prime}-A\hat a^{\prime 2}+ \alpha \right]
\label{eos1}
\end{equation}
where the following notation is used
\begin{equation}
\alpha\equiv \frac{1}{2r^2}\left[ \frac{\ddot T_m}{T_m}-\frac{\dot T_m^2}{T_m^2}\right]-\frac{1-2A}{2r}\left[ \frac{T_m^{\prime}}{T_m}-\frac{P^{\prime}}{P}\right]+\frac A2 \left[ \frac{T_m^{\prime\prime}}{T_m}-\frac{P^{\prime\prime}}{P}-\frac{T_m^{\prime2}}{T_m^2}+\frac{P^{\prime2}}{P^2}\right]
\label{tau}
\end{equation}

and the derivatives of the metric function $\hat a$ are given by the expressions in (\ref{redefi}), i.e., 
$
{\displaystyle \hat a^{\prime}=-x(r)+h^{\prime}=-x-\frac{\Sigma_+}{r\sqrt{2A}}}$, ${\displaystyle \dot {\hat a}=-\xi(\theta)-\frac{\Sigma_{-}}{\sqrt{2}}}
$.
After an easy but cumbersome calculation, the EoS relating the density $\rho$ of the source with  the pressure $T_m$ and anisotropies can be expressed as follows:
\begin{equation}
\rho =\frac{T_m}{8\pi P} \left[8 \pi \mu+ \alpha -2 A d(r,\theta)-2\frac{\Sigma_{+}^2}{r^2}+4 A x^2 \right]
\label{eosfinal}
\end{equation}

\section{Sample models}

As an example to illustrate the procedure, we provide  in this section specific expressions for the metric functions  as well as anisotropies of the source describing suitable models.

Firstly we need to solve the equation for the function $\Sigma_+$ along with its corresponding boundary conditions.
The boundary conditions established for $\mathbb{F}$ and $\mathbb{G}$ (see equations (\ref{Bcero}), (\ref{uncuartoequiv})) lead to the following behaviour for the function $\Sigma_{+}$:
\begin{equation}
\lim_{r \rightarrow 0} \Sigma_{+}=\frac{1}{\sqrt{2}}, \quad  \lim_{r \rightarrow 0} \Sigma_{+}^{\prime}=0,  \quad  \lim_{r \rightarrow r_{\Sigma}} \Sigma_{+}=\sqrt{L}
\label{sigma+behaves}
\end{equation}

 A suitable analytical solution for the function $\Sigma_{+}$ can be obtained by means of  the following expression
\begin{equation}
\Sigma_{+}=\frac{r\sqrt{A}}{2\sqrt{2}}J(r)-\frac{W}{2\sqrt{2}}
\label{6.4}
\end{equation}
where $W(r,y\equiv\cos\theta)$ is a solution of the following equation derived from (\ref{consistency})
\begin{equation}
\partial_y\left[(1-y^2)W_y\right]=r\sqrt{A}\partial_r\left[  \beta^{+} -\frac{W^2}{2}-r^2A \partial_r\left(\frac{W}{r\sqrt{A}} \right) \right] 
\label{6.5}
\end{equation}
with the corresponding boundary conditions derived from  (\ref{sigma+behaves}):
\begin{equation}
\lim_{r \rightarrow 0} W=\lim_{r \rightarrow 0} W_r=0,  \quad  \lim_{r \rightarrow r_{\Sigma}}W=\frac{1-A_{\Sigma}}{2\sqrt{A_{\Sigma}}}+4\sqrt{A_{\Sigma}}r_{\Sigma}\hat{\psi}^{\prime}_{\Sigma},
\label{Wbehaves}
\end{equation}
and 
\begin{equation}
\beta^{+}\equiv -8Ar^2\hat \beta+(A-1)\left[\frac 32 (A+1)-\frac{(A-1)}{(2Z)^2}-\frac{3\sqrt{A}}{2Z} \right]
\label{betaplus}
\end{equation}

An appropriate solution for $W$ can allways be obtained since the function $\beta^{+}$ contains the anisotropy $\Pi_{31}$ (see eq.(\ref{alfayotras})) which is  a function free of constraints. Therefore,  we propose a suitable function $W$ verifying the boundary conditions (\ref{Wbehaves}) as well as the required for satisfying (\ref{efeceroy1}) when the integration (\ref{efeintegrada}) were performed, and then the necessary anistropy $\Pi_{31}$ is obtained from (\ref{6.5}). Let us proceed now to calculate the  required function $W$ from the explicite integration of equation (\ref{efeintegrada}) with the function (\ref{6.4}), which leads to
\begin{equation}
\mathbb{F}=-a_E+\int (-x+h^{\prime}) \ dr=-a_E+V(r)+\frac 14\int \frac{W}{r\sqrt{A}}  \ dr
\label{6.7}
\end{equation}
with
\begin{equation}
V(r)=-\frac 34 \ln(2Z)+\frac r4 -\sqrt{\frac{\tau}{2}}\frac{r_{\Sigma}}{8}\ln\left( \frac{r_{\Sigma}+\sqrt{2/\tau} \ r}{r_{\Sigma}-\sqrt{2/\tau} \ r}\right),
\label{6.8V}
\end{equation}
 $\tau\equiv r_{\Sigma}/M$ being the compression parameter of the source, which is restricted to be $\tau>8/3$ by the strong energy condition $P<\mu$ (see \cite{weylsources} for details), and
hence, in addition to the boundary conditions (\ref{Wbehaves}),  $W$ must be a function satisfying
\begin{equation}
\lim_{r \rightarrow 0} \int \frac{W}{r\sqrt{A}} =-4 V_0,  \quad  \lim_{r \rightarrow r_{\Sigma}} \int \frac{W}{r\sqrt{A}}=4(\hat \psi_{\Sigma}-V_{\Sigma})
\label{Wintegral}
\end{equation}
where $V_0\equiv V(r=0)$ and $V_{\Sigma}\equiv V(r=r_{\Sigma})$.

The conditions (\ref{Wbehaves}), (\ref{Wintegral}) require that the   function $W$ be as follows:

\begin{equation}
W=r\sqrt{A} \left[ 4 \nu_1^{\prime} \hat \psi_{\Sigma}+4 \nu_2^{\prime} \hat \psi^{\prime}_{\Sigma}+\frac{24}{r_{\Sigma}}s(1-s)(V_0-V_{\Sigma})+\frac{1-A_{\Sigma}}{2 A_{\Sigma} r_{\Sigma}}s(3s-2)\right]
\label{Wdef}
\end{equation}

where the functions $\nu_1^{\prime}$ and $\nu_2^{\prime}$ are the derivatives of instrumental functions in the radial variable defined by the following behaviour:
\begin{equation}
\nu_1(r)=\left\lbrace
\begin{matrix}
& 0 & , \quad s=0  \nonumber \\
& 1 & , \quad s=1 
\end{matrix} 
\right.
\quad , \quad 
\nu_1^{\prime}(r)=\left\lbrace
\begin{matrix}
&0&, \quad s=0  \\
&0&, \quad s=1 
\end{matrix} 
\right.
\label{nu1}
\end{equation}
\begin{equation}
\nu_2(r)=\left\lbrace
\begin{matrix}
&0&, \quad s=0 \\
&0&, \quad s=1 
\end{matrix} 
\right.
\quad , \quad 
\nu_2^{\prime}(r)=\left\lbrace
\begin{matrix}
&0&, \quad s=0  \\
&1&, \quad s=1 
\end{matrix} 
\right.
\label{nu2}
\end{equation}

The calculation of these functions is addressed to the Appendix.

With the function $W$ in (\ref{Wdef}), the anistropy $\Pi_{31}$ is constrained by the equation (\ref{6.5}), since  $\beta^{+}$ for this model must be equal to

\begin{equation}
\beta^{+}(W)=4 \nu_1 \partial_y[(1-y^2)\partial_y \hat\psi_{\Sigma}]+ 4 \nu_2 \partial_y[(1-y^2)\partial_y \hat\psi^{\prime}_{\Sigma}]+\frac{W^2}{2}+r^2A \partial_r\left(\frac{W}{r\sqrt{A}}\right)
\label{betaplusW}
\end{equation}
and hence
\begin{equation}
\Pi_{31}=\frac{T_m}{4 \pi P}\left[A(\beta+d^+)-\frac{1}{8r^2}(A-1)\left[\frac 32 \frac{(A+1)}{A}-\frac{(A-1)}{2(2Z)^2}-\frac{3\sqrt{A}}{2Z} \right]+\frac{\beta^+(W)}{8r^2}\right]
\label{pi31}
\end{equation}

with the notation ${\displaystyle d(r,\theta)=-\frac{4\pi P}{A T_m}\Pi_{31}+ d^+}$ (see eq (\ref{alfayotras})).

In order to show a complete description of the model we shall calculate now the functions $\Lambda$ and $\Omega$ from (\ref{uves}) which allows us to determine the other anisotropies $\Pi_{23}$, $\Pi_{xy}$ (see the notation in eq. (\ref{lambdaomegadeltaybeta})):

\begin{eqnarray}
\Lambda&=& \frac{r^2A}{16} \left[ J- 4 \nu_1^{\prime} \hat \psi_{\Sigma}-4 \nu_2^{\prime} \hat \psi^{\prime}_{\Sigma}-\frac{24}{r_{\Sigma}}s(1-s)(V_0-V_{\Sigma})-\frac{1-A_{\Sigma}}{2A_{\Sigma}r_{\Sigma}}s(3s-2)\right]^2 +\nonumber\\
&-&\frac{(1-y^2)}{16} \left[(4 \nu_1^{\prime}\partial_y\hat \psi_{\Sigma}+4 \nu_2^{\prime} \partial_y\hat \psi^{\prime}_{\Sigma})^2 \right]\nonumber \\
\Omega&=& \frac{\sqrt{1-y^2}}{16} \left[4 \nu_1^{\prime}\partial_y\hat \psi_{\Sigma}+4 \nu_2^{\prime} \partial_y\hat \psi^{\prime}_{\Sigma} \right] \times \nonumber\\ 
&\times&\left[ J- 4 \nu_1^{\prime} \hat \psi_{\Sigma}-4 \nu_2^{\prime} \hat \psi^{\prime}_{\Sigma}-\frac{24}{r_{\Sigma}}s(1-s)(V_0-V_{\Sigma})-\frac{1-A_{\Sigma}}{2A_{\Sigma}r_{\Sigma}}s(3s-2)\right]
\label{lambdayomega}
\end{eqnarray}

And finally, the metric functions and the EoS have the following expressions:
\begin{eqnarray}
\hat a&=&V(r)+ \nu_1 \hat \psi_{\Sigma}+ \nu_2 \hat \psi^{\prime}_{\Sigma}-V_0+ \nonumber\\
&+&s^2\left[ (3-2s)(V_0-V_{\Sigma})+\frac{1-A_{\Sigma}}{2A_{\Sigma}}(s-1)\right]
\label{ahatfinal}
\end{eqnarray}
Let us note that this expression is the generalization of the special case $\mathbb{F}=0$ considered in \cite{weylsources}, where only the term\footnote{Indeed the simplest  type of functions for $\nu_1$ and $\nu_2$ was considered in that paper; the particular case $N=0$ for the instrumental functions was used (see eq.(\ref{nussol}) in Appendix)} $a_E= \nu_1 \hat \psi_{\Sigma}+ \nu_2 \hat \psi^{\prime}_{\Sigma}$  was used to define $\hat a$. In addition we observe that the dependence of $\hat a$ on the angular variable is the same and it is fixed by the exterior gravitational field. The other metric function $\hat g$ is obtained from (\ref{gvsa}) with a suitable pressure $T_m$ that fulfills the conditions (\ref{TmE})-(\ref{cero3}). Without limiting the generality, this pressure can be written as follows

\begin{eqnarray}
T_m&=&-\frac{M\mu}{r_{\Sigma}(r_{\Sigma}-2M)} e^{2\hat \Psi_{\Sigma}-2\hat\gamma_{\Sigma}}\left[(r-r_{\Sigma})+\left(\frac{1}{r_{\Sigma}}+4(\hat \Psi^{\prime}_{\Sigma}-\hat \gamma^{\prime}_{\Sigma}) \right)\frac{(r-r_{\Sigma})^2}{2} \right]\nonumber\\
&+&(r-r_{\Sigma})^3 \hat T(r,\theta)
\label{ejemTm}
\end{eqnarray}

where $\hat T(r,\theta)$ is  an arbitrary function such that $T_m$ satisfies the conditions\footnote{For example, a polynomic function in the radial variable $r$, $\hat T=\hat t_0(\theta)+\hat t_1(\theta) r+\hat t_2(\theta)r^2$ can be used and the functions $\hat t_i$ are determined from those conditions at the origin.} at $r=0$ (\ref{cero1})-(\ref{cero3}).

We obtain the EoS of this model by putting into the equation (\ref{eosfinal})  the particular expressions for the functions $\Sigma_+$ (\ref{6.4}) and $d(r,\theta)$ which depends on $\beta^+(W)$ (\ref{betaplusW}):

\begin{eqnarray}
\rho &=&\frac{T_m}{8\pi P} \left[8 \pi \mu+ \alpha +4 A x^2 +2A\beta-\frac 14 A J^2+\frac{\sqrt{A}}{2r}J W-\frac 18 \frac{W^2}{r^2}+\right. \nonumber\\
&+&\left.\frac{1}{r^2}\left[ \nu_1 \partial_y[(1-y^2)\partial_y \hat\psi_{\Sigma}]+  \nu_2 \partial_y[(1-y^2)\partial_y \hat\psi^{\prime}_{\Sigma}]\right]\right.+\nonumber\\
&+&\left.\frac  A4\partial_r\left(\frac{W}{r\sqrt{A}} \right)-\frac{(A-1)}{4r^2}\left[\frac 32 \frac{(A+1)}{A}-\frac{(A-1)}{2(2Z)^2}-\frac{3\sqrt{A}}{2Z} \right]\right]
\label{eosmodel}
\end{eqnarray}
which after simplification becomes
\begin{eqnarray}
\rho &=&\frac{T_m}{8\pi P} \left[8 \pi \mu+ \alpha +\frac  A4\partial_r\left(\frac{W}{r\sqrt{A}} \right)+\frac{\sqrt{A}}{2r}J W-\frac 18 \frac{W^2}{r^2}+\right.\nonumber\\
&+&\left.\frac{1}{r^2}\left[ \nu_1 \partial_y[(1-y^2)\partial_y \hat\psi_{\Sigma}]+  \nu_2 \partial_y[(1-y^2)\partial_y \hat\psi^{\prime}_{\Sigma}]\right]\right.+\nonumber\\
&+&\left.\frac{1}{8r^2}\left[ \frac{(-9A^2+4A+1)}{A}+\frac{3(A-1)^2}{(2Z)^2}-\frac{2(1-A)(17A-2)}{(2Z)\sqrt{A}} \right]\right]
\label{eosmodel}
\end{eqnarray}

\section{Conclusions}

The exterior gravitational field of a compact object constrains the physics on the inside of the source. Once  a Weyl vacuum solution  describing the gravitation of an axially static stelar object is approppriately matched with an interior metric then both  the line element as well as the energy-momentum tensor of the source can be constructed in terms of the exterior metric functions. 
A global description of the gravitation  allows us to relate the energy-momentun tensor with the exterior gravitational field.

In this paper we have contributed with good behavioured models of sources, both in the spherical and non-spherical cases, providing suitable EoS and the global metric by means of the knowledge as a starting point of any vacuum solution of the Weyl family. In these models it is shown how the different exterior metrics contribute in a different way into the anisotropies or the density of the source, at the same time that the pressure $T_m$ depends on those exterior metric functions. How the pressures and the EoS undertake the presence of non-spherical symmetry at the exterior, being led by the departure of the exterior metric functions $\hat \psi_{\Sigma}$, $\hat \gamma_{\Sigma}$ from the corresponding Schwarzschild ones $\psi_{\Sigma}^s$, $\gamma_{\Sigma}^s$,  is shown.
Therefore, we get information about the physics on the inside of the source from the behaviour of the exterior gravitational field.

These results do not establish, of course, general relationships since we have been dealing with the so called Weyl gauge, and we manage only two metric functions in the interior metric; that condition on the Einstein tensor is allways satisfied on the vacuum, however for the interior space-time it implies a restriction on possible solutions. Nevertheless, the results presented here show that, within that restriction on possible solutions, the matter distribution of the source assumes some constraints derived from the exterior gravitational field of the global metric.
 These constraints can be expressed in terms of the gravitational field which is matching the interior solution. Since the method proposed \cite{weylsources} allows us to constract a well matched interior metric for any exterior solution of the Weyl family, we can obtain for any vacuum solution how  the energy-momentun tensor of axially symmetric static  sources are affected by different physical characteristics of the gravitional field.

And last but not least, we have introduced an expression derived from the matching conditions that allows us to calculate the RMM in terms of the source. An example of that is obtained for the quadrupole moment of the metric which is related, by means of a source integral (\ref{Q}), with the physics of the source (anisotropies and pressures).

\section*{Acknowledgments}
This  work  was  partially supported by the Spanish {\it Ministerio de Ciencia, Innovaci\'on y Universidades} under Research Project reference PGC2018-096038-B-I00, as well as the Consejer\'\i a
de Educaci\'on of the Junta de Castilla y Le\'on under the Research Project Grupo de Excelencia GR234.

\section{Appendix}

\subsection{The boundary and regularity conditions}  The conditions (\ref{boundarycond}) for the function  $\mathbb{G}$ and its first derivative shall be satisfied if we still hold the behaviour at the origin for $T_m$ and its derivative as in the spherical previous case, i.e. $T_m(s=0)=P(s=0)$, $T_m^{\prime}(s=0)=P^{\prime}(s=0)$. This conclusion can be inferred from equation (\ref{gvsa}):
\begin{equation}
\ln\left( \frac{T_m}{P}\right)=2\hat a-2\hat g \quad , \quad \left[ \frac{T_m^{\prime}}{T_m}-\frac{P^{\prime}}{P}\right]=2\hat a^{\prime}-2\hat g^{\prime}
\label{Fprima}
\end{equation}
But at the boundary $r=r_{\Sigma}$ ($s=1$), the pressure $T_m$ has to take into account the exterior gravitational field as follows
\begin{eqnarray}
\ln\left( \frac{T_m}{P}\right)\mid_{\Sigma}&=&2\hat \psi_{\Sigma}-2\hat \gamma_{\Sigma} \ , \ \left[ \frac{T_m^{\prime}}{T_m}-\frac{P^{\prime}}{P}\right]\mid_{\Sigma}=2\hat \psi^{\prime}_{\Sigma}-2\hat \gamma^{\prime}_{\Sigma} \nonumber\\
\left[ \frac{\dot T_m}{T_m}\right]\mid_{\Sigma}&=&2\dot{\hat \psi}_{\Sigma}-2\dot{\hat \gamma}_{\Sigma}
\label{FprimaE}
\end{eqnarray}
Since $P(s=1)=0$ we have to avoid the divergencies in equation (\ref{FprimaE}) by imposing $T_m(s=1)=0$ as well as the following conditions on the first and second derivatives of $T_m$ at the boundary arising from the resolution of the indeterminations in (\ref{FprimaE}):

\begin{eqnarray}
\lim_{r \rightarrow r_{\Sigma}}T_m^{\prime}&=&P^{\prime}_{\Sigma} e^{2\hat \Psi_{\Sigma}-2\hat \gamma_{\Sigma}}=-\frac{M\mu}{r_{\Sigma}(r_{\Sigma}-2M)} e^{2\hat \Psi_{\Sigma}-2\hat\gamma_{\Sigma}}\nonumber\\
\lim_{r \rightarrow r_{\Sigma}}T_m^{\prime\prime}&=&T_m^{\prime}|_{\Sigma}\left[\frac{P^{\prime\prime}_{\Sigma}}{P^{\prime}_{\Sigma}}+4\left(\hat \Psi^{\prime}_{\Sigma}-\hat \gamma^{\prime}_{\Sigma} \right) \right]=T_m^{\prime}|_{\Sigma}\left[\frac{1}{r_{\Sigma}}+4\left(\hat \Psi^{\prime}_{\Sigma}-\hat \gamma^{\prime}_{\Sigma} \right) \right]\nonumber\\
\lim_{r \rightarrow r_{\Sigma}}\dot T_m&=&0 \nonumber\\
\lim_{r \rightarrow r_{\Sigma}}\dot T_m^{\prime}&=&T_m^{\prime}|_{\Sigma}\left(2\dot{\hat \psi}_{\Sigma}-2\dot{\hat \gamma}_{\Sigma}\right)=-\frac{M\mu}{r_{\Sigma}(r_{\Sigma}-2M)} e^{2\hat \Psi_{\Sigma}-2\hat\gamma_{\Sigma}}\left(2\dot{\hat \psi}_{\Sigma}-2\dot{\hat \gamma}_{\Sigma}\right)\nonumber \\
\label{TmprimasenE}
\end{eqnarray}

Hence, the required remaining boundary conditions have to be with the second derivative of $\mathbb{G}$ as well as the behaviour of $\mathbb{F}$ and its first derivative in the extremes of the interval $s \in [0,1]$:

\vskip 2mm

{\bf A)} Let us start with $\mathbb{F}^{\prime}$. The requirement $\mathbb{F}^{\prime}(s=0)=0$ forces to choose the negative global sign in  (\ref{H}) in order to  get that condition without any divergence at the origin, since $\mathbb{F}^{\prime}$ is

\begin{equation}
\mathbb{F}^{\prime}=-a_{E}^{\prime}+\frac{1}{2r \sqrt{A}}\left[\frac{3\sqrt{AA_{\Sigma}}-1}{2Z} - \sqrt{2\Lambda + 2 \sqrt{\Lambda^2+4Ar^2\Omega^2}}\right]\equiv -a_{E}^{\prime}+\frac{B}{2r \sqrt{A}}
\label{h}
\end{equation}

On the one hand, from (\ref{h}) the condition ${\displaystyle \lim_{r \rightarrow 0} \mathbb{F}^{\prime}=0}$  needs  the function $B$ to be an infinitesimal of order $O(r^2)$ at the origin, or equivalently  that  implies the following limits
\begin{equation}
B_0\equiv \lim_{r \rightarrow 0}B =0 , \quad B_0^{\prime}\equiv \lim_{r \rightarrow 0} B^{\prime}=0
\label{fprimacero}
\end{equation}
These limits  $B_0=B_0^{\prime}=0$  imply the following conditions for the functions $\Lambda$ and $\Omega$ at the origin, (where the subindex denotes the value at the origin)
\begin{equation}
\Lambda_0+\sqrt{\Lambda^2+4Ar^2\Omega^2}|_0=\frac 12, \quad \left(\sqrt{\Lambda+\sqrt{S}}\right)^{\prime}_0=0
\label{Bcero}
\end{equation} and we can see from (\ref{lambdaomegadeltaybeta}) that $\Lambda_0=\delta_0-\xi^2+\frac 14$  and then equation (\ref{Bcero}) is equivalent to
\begin{equation}
\delta_0-\xi^2+(4Ar^2\Omega^2)\mid_0=O(r^2)_0,
\label{deltacero}
\end{equation}
but, the value at the origin of the last term is an infinitesimal of order $O(r^2)$ iff also ${\displaystyle \frac{\partial_{\theta}T_m}{T_m}}$  is so, and hence in addition  $(4Ar^2\Omega^2)\mid_0=\xi^2$. With this condition  on the partial derivative of $T_m$ the second limit of equation (\ref{fprimacero}) is also verified, so we conclude that the boundary condition for $\mathbb{F}^{\prime}$ at the origin is fulfilled if $\delta_0$ vanishes behaving as an infinitesimal of order $O(r^2)$:
\begin{equation}
\delta_0=-\xi\frac{\partial_{\theta}T_m}{T_m} \mid_0=0,
\label{deltaceroimply}
\end{equation}

On the other hand, from (\ref{h}) the condition ${\displaystyle \lim_{r \rightarrow r_{\Sigma}} \mathbb{F}^{\prime}=0}$ leads to
\begin{equation}
L\equiv \frac{(3A_{\Sigma}-1-4A_{\Sigma}r_{\Sigma}\hat\psi^{\prime}_{\Sigma})^2}{8A_{\Sigma}} =\lim_{r \rightarrow r_{\Sigma}} \Lambda+\sqrt{\Lambda^2+4Ar^2\Omega^2}
\label{uncuartoequiv}
\end{equation}
This condition (\ref{uncuartoequiv}) becomes an identity if one takes into account the behaviour at the boundary of the functions $\Lambda$ and $\Omega$ from the equations (\ref{lambdaomegadeltaybeta}). Let us note that $\Pi_{xy}|_{\Sigma}=0$ as well as $\Pi_{31}+\Pi_{23}|_{\Sigma}=0$ since $T_1^1|_{\Sigma}=T_2^2|_{\Sigma}=0$ because  the pressure $P$, and both  $\hat p_{xy}$ and $\hat p_{xx}$ vanish at the boundary (the proof is given in \cite{weylsources}). After a cumbersome calculation we obtain that this equation  (\ref{uncuartoequiv}) is ever satisfied for those functions\footnote{As an example, the specific case of matching with the exterior Schwarzschild metric is easy to verify since $\hat{\psi}_{\Sigma}=\hat \gamma_{\Sigma}=0$, and hence $\alpha_{\Sigma}=\delta_{\Sigma}=0$. In the general case we need to make use of the derivatives of the function $\hat \gamma$ which are functions of $\hat\psi$ from the vacuum field equations (see \cite{weylsources}  for details.)
}
\vskip 2mm

{\bf B)} With respect to the behaviour of $\mathbb{F}$ ($\mathbb{F}(s=0)=\mathbb{F}(s=1)=0$) we deduce from  (\ref{efeintegrada}) that 
\begin{equation}
\hat\psi_{\Sigma}=\int_0^{r_{\Sigma}} (-x+h^{\prime}) dr
\label{efeceroy1}
\end{equation}
We comment on the relevance of this expression at the section V.

\vskip 2mm

{\bf C)} Finally, the second derivative of the function  $\mathbb{G}$ must be null at the origin. From  equation (\ref{gvsa}) we have  that
\begin{equation}
\left[ \frac{T_m^{\prime\prime}}{T_m}-\frac{P^{\prime\prime}}{P}
-\left(\frac{T_m^{\prime}}{T_m}\right)^2
+\left(\frac{P^{\prime}}{P}\right)^2\right]=2\hat a^{\prime\prime}-2\hat g^{\prime\prime}
\label{G2prima}
\end{equation}
Since we have assumed that $T_m(s=0)=P(s=0)$, $T_m^{\prime}(s=0)=P^{\prime}(s=0)$ then $\mathbb{G}^{\prime\prime}(s=0)=0$   impose the following condition (take into account equations ({\ref{efeintegrada}), (\ref{aygsimple})):
	\begin{equation}
	\left[ \frac{T_m^{\prime\prime}}{T_m}-\frac{P^{\prime\prime}}{P}
	\right]_0=-2x^{\prime}_0+2h^{\prime\prime}_0 =2 \lim_{r\rightarrow 0}\partial_r \left[-x+h^{\prime}\right]
	\label{G2primacero}
	\end{equation}
	In order to evaluate the behaviour of $-x^{\prime}+h^{\prime\prime}$ at the origin, we  solve the above limit as followws:
	\begin{equation}
	\lim_{r\rightarrow 0}\partial_r \left[ \frac{B}{2 r\sqrt{A}}\right]=\frac 14 B^{\prime\prime}_0,
	\label{limitcero}
	\end{equation}
	where we have taken into account that $B$ is an infinitesimal of order $O(r^2)$ (as we already impossed before). Consequently, the equation (\ref{G2primacero}) turns out to be
	\begin{equation}
	\left[ \frac{T_m^{\prime\prime}}{T_m}-\frac{P^{\prime\prime}}{P}
	\right]_0=\frac 12 B^{\prime\prime}_0,
	\label{Tm2primacero}
	\end{equation}
	The evaluation of the limit at the origin of the second derivative of $B$ needs a large amount of patience, but finally we solve that
	\begin{equation}
	B^{\prime\prime}_0= -\Lambda^{\prime \prime}_0-2 S^{\prime \prime}_0 \sin^2\theta  -\frac{2M}{r_{\Sigma}^3} \frac{3\sqrt{A_{\Sigma}}+1}{3\sqrt{A_{\Sigma}}-1}
	\label{B20previo}
	\end{equation}
	and finally the evaluation of $\Lambda^{\prime \prime}_0$ and $S^{\prime \prime}_0$ leads to a relation between the anisotropies $\Pi_{31}+\Pi_{32}$ and the pressure $T_m$ as follows:
	\begin{equation} 8\pi(\Pi_{31}+\Pi_{32})_0=-\frac{3\sqrt{A_{\Sigma}}}{3\sqrt{A_{\Sigma}}-1}\frac{\cos\theta}{\sin\theta}\partial_r^2\left(\frac{\partial_{\theta}T_m}{T_m} \right)|_0 
	\label{anis0}
	\end{equation}

	\subsection{The integrability condition}
	
	Both equations (\ref{cuadratura1}), (\ref{cuadratura2}) have to satisfy the corresponding integrability condition ${\displaystyle \partial_{\theta}(H)=\partial_r\left(\frac{\Omega}{H}\right)}$, leading to the following equation:
	\begin{equation}
	\partial_{\theta}\left( \Lambda+\sqrt{S} \right)=4 A r^2 \left[ \Omega^{\prime}-\Omega\left(\frac{(\Lambda+\sqrt{S})^{\prime}}{2(\Lambda+\sqrt{S})}+\frac{1-2A}{rA} \right)\right]
	\label{ic1}
	\end{equation}
	
	If we make use of the relation $(\Lambda+\sqrt{S})(-\Lambda+\sqrt{S})=4Ar^2\Omega^2$ then another but equivalent equation can be obtained  since the following equation for $\dot y$ holds
	\begin{equation}
	\dot y =\frac{\Omega}{H}=-\frac{1}{\sqrt 2}\sqrt{-\Lambda+\sqrt{S}},
	\end{equation}
	and then the integrability condition (\ref{ic1}) leads to the following equation
	\begin{equation}
	\partial_{\theta}\left( \Lambda+\sqrt{S}\right)=2Ar^2 \Omega \left[\frac{(-\Lambda+\sqrt{S})^{\prime}}{-\Lambda+\sqrt{S}}\right]
	\label{ic2}
	\end{equation}
	and consequently,
	\begin{equation}
	\partial_{\theta}\left( \sqrt{\Lambda+\sqrt{S}}\right)=r \sqrt{A}\partial_{r}\left(\sqrt{-\Lambda+\sqrt{S}}\right)
	\label{ic3}
	\end{equation}
	
\vskip 2mm
	
	\subsection{The instrumental functions $\nu_1$, $\nu_2$}
	
	The functions  $\nu_1$, $\nu_2$ are introduced (\ref{nu2}) to construct a suitable function $W$ compatible with the boundary conditions. Both of them are functions of second order in their variable, the radial coordinate, nearby  the origin. It means  that the function and its derivative must vanish at the origin. At the same time we are forced to choose these functions in such a way that they behaves in a particular way at the boundary of the source ($r=r_{\Sigma}$, or $s\equiv r/ r_{\Sigma}=1$): the function $\nu_1$ takes the value $1$ and its derivative vanishes at that point, whereas the function $\nu_2$ vanishes when $s=1$ and its derivatives becomes $1$ at the boundary.
	
	Without any lack of generality\footnote{These instrumental functions can be polynomials or whatever other function with the Taylor series at the origin given by the polynomials obtained here.}, these conditions are satisfied by the following polynomials of arbitrary degree
	\begin{equation}
	\nu_{k}(r)=\sum_{j=2} b_j^{(k)} s^j
	\label{polinomio}
	\end{equation}
	with the following constraints
	\begin{eqnarray}
	\sum_{j=2} b_j^{(k)} = \delta_{k1} \nonumber \\
	\sum_{j=2} j b_j^{(k)} =  r_{\Sigma} \ \delta_{k2}
	\label{condpolis}
	\end{eqnarray}
	where $k$ takes the value $k=1,2$ for the corresponding function $\nu_1$ or $\nu_2$ respectively.
	
	The solution of these indeterminated compatible linear system of algebraic equations leads to the following functions 
	\begin{eqnarray}
	\nu_1(r)&=&s^2\left[ 3-2s+\sum_{i=0}^N n_i^{(1)} (1+i-s(i+2)+s^{i+2})\right] \nonumber \\
	\nu_2(r)&=&s^2\left[ r_{\Sigma}  (s-1)+\sum_{i=0}^N n_i^{(2)} (1+i-s(i+2)+s^{i+2})\right] \nonumber\\
	\label{nussol}
	\end{eqnarray}
	where $n_i^{(k)}$ are arbitrary coeficients.

\end{document}